\documentclass{article}
\usepackage{arxiv}

\usepackage[utf8]{inputenc} 
\usepackage[T1]{fontenc}    
\usepackage{hyperref}       
\usepackage{url}            
\usepackage{booktabs}       
\usepackage{amsfonts}       
\usepackage{amsmath}
\usepackage{nicefrac}       
\usepackage{microtype}      
\usepackage{lipsum}		
\usepackage{graphicx}
\usepackage{natbib}
\usepackage{doi}
\usepackage{footmisc}
\usepackage{bbold}
\usepackage{listings}
\usepackage{threeparttable}

\PassOptionsToPackage{numbers, compress}{natbib}

\title{Estimands for Randomized Discontinuation Designs in Oncology}

\author{ \href{https://orcid.org/0000-0001-5461-1737}{\includegraphics[scale=0.06]{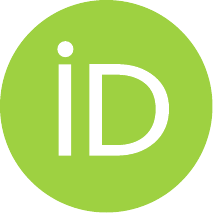}\hspace{1mm}Ayon Mukherjee}\\
	Population Health Sciences Institute\\
	Newcastle University\\
	Newcastle, United Kingdom \\
	\texttt{ayon.mukherjee@newcastle.ac.uk} \\
	\And
	\href{https://orcid.org/0000-0002-1626-2588}{\includegraphics[scale=0.06]{orcid.pdf}\hspace{1mm}Oleksandr Sverdlov} \\
    Advanced Quantitative Scieces\\
	Novartis Pharmaceuticals Corporation\\
	East Hannover, New Jersey, USA\\
	\texttt{alex.sverdlov@novartis.com} \\
    \And
	\href{https://orcid.org/0000-0002-4846-7238}{\includegraphics[scale=0.06]{orcid.pdf}\hspace{1mm}Ngoc-Thuy Ha} \\
    Biostatistics Oncology Late Phase Development\\
	Merck KGaA\\
	Darmstadt, Germany\\
	\texttt{ngoc-thuy.ha@merckgroup.com} \\
    \And
	\href{https://orcid.org/0009-0004-4554-499X}{\includegraphics[scale=0.06]
    {orcid.pdf}\hspace{1mm}Yu Deng} \\
    Biostatistics \\
	Genentech Inc.\\
	South San Francisco, California, USA\\
	\texttt{deng.yu@gene.com} \\
}

\renewcommand{\shorttitle}{\textit{arXiv} Template}

\hypersetup{
pdftitle={A template for the arxiv style},
pdfsubject={q-bio.NC, q-bio.QM},
pdfauthor={David S.~Hippocampus, Elias D.~Striatum},
pdfkeywords={First keyword, Second keyword, More},
}

\begin{document}
\maketitle

\begin{abstract}
Randomized discontinuation design (RDD) is an enrichment strategy commonly used to address limitations of traditional placebo-controlled trials, particularly the ethical concern of prolonged placebo exposure.  RDD consists of two phases: an initial open-label phase in which all eligible patients receive the investigational medicinal product (IMP), followed by a double-blind phase in which responders are randomized to continue with the IMP or switch to placebo. This design tests whether the IMP provides benefit beyond the placebo effect. The estimand framework introduced in ICH~E9(R1) strengthens the dialogue among clinical research stakeholders by clarifying trial objectives and aligning them with appropriate statistical analyses. However, its application in oncology trials using RDD remains unclear. This manuscript uses the phase~III  JAVELIN Gastric~100 trial and the phase~II trial of sorafenib (BAY 43-9006) as case studies to propose an estimand framework tailored for oncology trials employing RDD in phase~III and phase~II settings, respectively.  We highlight some similarities and differences between RDDs and traditional randomized controlled trials in the context of ICH~E9(R1). This approach aims to support more efficient regulatory decision-making.
\end{abstract}


\section{Introduction}
Randomized discontinuation design (RDD) (also known as ``randomized withdrawal design'' or ``randomized discontinuation trial design'') is occasionally used in clinical trials testing the long-term effectiveness of non-curative therapies, where the long-term use of placebo cannot be justified on ethical or practical grounds \citep{AmeryDony1975, Kopec1993}. In the original version of RDD \citep{AmeryDony1975}, the trial is divided into two periods, the open-label period during which all eligible participants are exposed to the experimental treatment, and the randomized period, during which only a subgroup of patients who responded to the experimental treatment (``responders'') are randomized in a double-blind manner to continue the experimental treatment or switch to placebo. The primary analysis tests the treatment's superiority to placebo in the selected subpopulation. According to the 2019 US FDA guidance on enrichment strategies for clinical trials
\citep{FDA2019}, RDD is classified as a predictive enrichment strategy, where the trial is enriched with apparent responders. The same guidance describes some successful use cases of RDD and mentions that this design ``has become the standard for studies to demonstrate the ability of psychotropic drugs to prevent recurrent depression, psychosis, and anxiety.''

RDD has also gained traction in phase~II oncology studies. \cite{Rosner2002} conceptualized the RDD for a phase~II oncology trial to evaluate the clinical activity of a cytostatic agent, acknowledging the heterogeneity of tumor growth rates in the population. The design initially treats all participants with the study drug and then randomizes only those  whose disease is ``stable'' to the study drug or placebo. Such a design allows to determine whether the disease stabilization is due to the drug or due to the selection of patients with naturally slow-growing tumors. \cite{Trippa2012} proposed a Bayesian decision-theoretic approach to optimize the parameters of a phase~II oncology RDD. Some successful applications of RDD in phase~II oncology studies include sorafenib in patients with advanced renal clear cell carcinoma \citep{Ratain2006}, tivozanib in patients with metastatic renal cell carcinoma \citep{Nosov2012}, and cabozantinib in patients with advanced prostate cancer \citep{Smith2013}, amongst others.

The use of RDD in phase III clinical trials is less common. Typically phase~III trials aim to provide definitive evidence of the efficacy and safety of the study treatment and utilize the traditional randomized controlled trial (RCT) design. Two recent applications of RDD in phase~III are the study of pegvaliase for the treatment of phenylketonuria (PKU) \citep{Harding2018} and the study of avelumab in patients with gastric cancers \citep{Moehler2020}.

The estimand framework, as outlined in the International Council for Harmonisation of Technical Requirements for Pharmaceuticals for Human Use (ICH) E9(R1) addendum (called E9(R1) hereafter), provides clarity in clinical trial objectives by aligning the targeted estimand with statistical analysis methods \citep{FDA2021}. Despite its benefits and wide applications in RCTs, to the best of our knowledge, the estimand framework has not been applied in oncology trials using RDD. This gap may be due to the lack of familiarity among investigators with intercurrent events (ICEs) that affect the observation or interpretation of endpoints in RDD. Developing estimands for RDD in oncology trials can enhance decision-making efficiency and provide a structured approach to handling ICEs, ultimately benefiting the drug registration process. Implementing this framework can lead to more robust and transparent trial outcomes, facilitating regulatory approval. 

In this paper, we propose an application of the estimand framework for oncology clinical trials designed with RDD. Section~\ref{Sec2} presents two motivating examples that implemented RDD without considering estimands as per E9(R1). Section~\ref{Sec3} describes the relevant scientific questions of interest and the estimand attributes, highlighting some similarities and differences between RDDs and the traditional RCTs. Section~\ref{Sec4} provides a discussion and some recommendations for practice.

\section{Motivating examples}\label{Sec2}

\subsection{Phase II trial of sorafenib}\label{Sec2.1}
Sorafenib (BAY 43-9006) is an oral multi-kinase inhibitor that works as a cytostatic agent by stabilizing tumor progression but not reversing the existing tumor. \cite{Ratain2006} reported the results of a phase~II trial using RDD to assess the activity of sorafenib in patients with advanced renal cell carcinoma (RCC). This study enrolled 202 RCC patients with histologically or cytologically confirmed metastatic refractory cancer, who received initial 12-week open-label treatment with oral doses of sorafenib at 400~mg twice daily. A total of 187 patients (93\%) completed the open-label phase. For these patients, individual disease status at week~12 was assessed based on WHO criteria (change in bidimensional tumor measurement from baseline), and the patients were categorized into three groups: those with $\ge$~25\% tumor shrinkage ($N$~=~79) were considered ``responders'' and continued taking sorafenib until clinical toxicity or disease progression; those with $\ge$ 25\% tumor growth ($N$~=~43) were considered having a ``progressive disease'' and were discontinued from the trial; and those with the change in tumor size less than $\pm$ 25\% ($N$~=~65) were considered having ``stable disease'' and moved into the double-blind part of the study, where they were randomized to sorafenib ($n$~=~32) or matching placebo ($n$~=~33).

The primary efficacy endpoint was the proportion of randomized patients who were progression-free at 12 weeks post-randomization (24~weeks from study entry). This proportion was 50\% for sorafenib-treated patients vs. 18\% for the placebo group ($p$ = 0.0077). The secondary endpoint, progression-free survival (PFS) from 12-week randomization was  statistically longer in the sorafenib group (median PFS~=~24 weeks) than in the placebo group (median PFS~=~6 weeks; $p$~=~0.0087). It was concluded that ``sorafenib has significant disease-stabilizing activity in metastatic renal cell carcinoma and is tolerable with chronic daily therapy.''  Following this successful phase~II study, there was the definitive phase~III trial \citep{Escudier2007} which confirmed the efficacy of sorafenib in patients with RCC and led to the FDA approval of the drug.

\subsection{Phase III JAVELIN Gastric 100 trial} \label{Sec2.2}

JAVELIN Gastric 100 trial \citep{Moehler2020} was an open-label, multicenter, randomized phase~III study in patients with advanced, human epidermal growth factor receptor~2 (HER2)-negative gastric cancer (GC) or gastroesophageal junction cancer (GEJC). All eligible participants had an open-label treatment with first-line induction chemotherapy for up to 12 weeks. Patients without progressive disease were randomized 1:1 to either maintenance therapy with avelumab 10mg/kg IV every 2 weeks or with the same chemotherapy.  The randomization was stratified by region (Asia vs. non-Asia). The primary endpoint was overall survival (OS) defined as time from randomization to death from any cause. Secondary endpoints included PFS, best overall response, duration of response, and safety. The primary objective was to test the superiority of avelumab maintenance therapy over the continuation of first-line chemotherapy with respect to OS in all randomized patients or in a subset of all randomized patients with PD-L1-positive tumors. 

In the induction phase of the trial, a total of 805 patients were enrolled and 799 patients received induction chemotherapy. Among them, 499 patients met the criteria of stable disease and were randomized to avelumab maintenance arm ($n$~=~249) or continued chemotherapy arm ($n$~=~250). The numbers of PD-L1-positive patients in the two randomized groups were 30 (avelumab maintenance arm) and 24 (continued chemotherapy arm). It was found that OS was not significantly different between the two arms in all randomized patients (stratified hazard ratio (HR) of avelumab vs. chemotherapy  was 0.91; 95\% CI, 0.74 to 1.11; stratified one-sided $p$~=~0.1779), nor in the pre-specified PD-L1 positive subgroup (stratified HR~=~1.13; 95\% CI, 0.57 to 2.22; stratified one-sided $p$~=~0.6352). At the same time, exploratory subgroup analyses showed some evidence of OS difference between the arms, favoring avelumab. Overall, it was concluded that JAVELIN Gastric 100 trial ``did not achieve its primary objective of OS improvement with maintenance avelumab in patients with disease control after induction chemotherapy for advanced GC/GEJC.''

\section{Estimands for RDD Trials}\label{Sec3}

Table~1 provides a comparative summary of the two trials described in Section~\ref{Sec2} in terms of the estimand attributes. For each attribute, we also highlight the differences between RDD and the traditional RCT.
\begin{table}[t]
\footnotesize
\begin{threeparttable}\label{Table1}
    \centering
    \caption{Applying the estimand framework to two oncology trials implemented with a randomized discontinuation design.}
    \begin{tabular}{p{1.3cm}p{3.8cm}p{3cm}p{3.8cm}p{3cm}}
    \hline
     & \textbf{Sorafenib phase~II trial \citep{Ratain2006}} & \textbf{What is the difference compared to the RCT?} & \textbf{JAVELIN Gastric 100 phase~III trial \citep{Moehler2020}} & \textbf{What is the difference compared to the RCT?} \\
    \hline
    \textbf{Objective}  & To assess whether maintenance therapy with sorafenib stabilizes disease compared to placebo in advanced RCC patients  with changes in bidimensional tumor measurements that were less than 25\% from baseline after 12 weeks of sorafenib & Similiar to RCT
    & To assess whether avelumab maintenance therapy improves OS compared to continued chemotherapy in advanced HER2-negative GC/GEJC patients without PD after up to 12 weeks of first-line chemotherapy with oxaliplatin plus fluoropyrimidine & Similar to RCT 
    \\
    \hline
    \textbf{Population} & Patients with advanced RCC who achieved SD (change in tumor size less than $\pm$25\%) after 12 weeks of sorafenib treatment & 1.~The definition of a population is dependent on the results of Part~1 of the study. $\quad\quad\quad\quad\quad\quad\quad\quad$ 2.~As sorafenib was experimental, the population beyond the study did not exist at the time when the trial was conducted. & Patients with advanced HER2-negative GC/GEJC who had no PD per RECIST~v1.1 after 12 weeks of chemotherapy & Same as \#1 for sorafenib \\
    \hline
    \textbf{Treatment} & Sorafenib vs. placebo & Placebo was the intervention for Part~2; however, the research hypothesis tested the superiority of sorafenib vs. placebo & Avelumab vs. chemotherapy & Similar to RCT \\
    \hline
    \textbf{Endpoint} & Progression status (Yes/No) at 12 weeks following random assignment & Binary primary endpoint was similar (albeit not identical) to the binary enrichment endpoint  & OS (time from randomization to death from any cause) & Similar to RCT \\
    \hline
    \textbf{Population-level summary} & Proportion of patients with progression-free status at 12 weeks following random assignment & Similar to RCT  & Median survival time following random assignment & Similar to RCT  \\ 
    \hline
    \textbf{ICEs} & Premature treatment discontinuation, with or without subsequent follow-up therapy -- treatment policy strategy & Similar to RCT & Premature treatment discontinuation, with or without subsequent follow-up therapy -- treatment policy strategy & Similar to RCT \\
    \hline
    \end{tabular}
    \begin{tablenotes}
    \footnotesize
       \item [] Abbreviations: GC = gastric cancer; GEJC = gastroesophageal junction cancer; HER2 = human epidermal growth factor receptor~2; ICEs = intercurrent events; OS~=~overall survival; PD = progressive disease; RCC~=~renal cell carcinoma; RDD = randomized discontinuation design; RCT = randomized controlled trial; SD = stable disease.
    \end{tablenotes}    
    \end{threeparttable}
\end{table}

As per E9(R1), the study objective refers to the overarching scientific question that the trial is designed to answer. In the phase~II trial example \citep{Ratain2006}, the study aimed at assessing the continued effect of sorafenib in patients who achieved stable disease after initial treatment with sorafenib, whereas the phase~III trial \citep{Moehler2020} aimed at assessing the efficacy of avelumab as a maintenance therapy in patients with stable disease initially treated with the standard 1st line chemotherapy. E9(R1) emphasizes that trial objectives must be clear and precise so they can be translated into key clinical questions of interest by defining appropriate estimands. In RDD, we must clearly outline the maintenance treatment period following randomization.

The population attribute is a challenging aspect of RDD because the ``response'' to treatment in the open-label period is used to select a subpopulation for the randomized period. \cite{Mutze2025} introduced four principles for a well-defined estimand, one of which is that the estimand should be independent of trial conduct and data. However, in RDD, the randomized population is retrospectively determined based on individual treatment responses - information that only becomes available during the trial. This makes the estimand dependent on a latent responder population, which cannot be identified in advance and may differ across studies or real-world settings. As a result, this data-driven population definition limits generalizability and challenges the estimand framework, which assumes a clearly defined target population before data collection.  An ``enriched'' population must be objectively defined. \cite{Fedorov2022} highlighted that a high rate of misclassifications (e.g., a ``true'' responder is classified as a non-responder or a ``true'' non-responder is classified as a responder) can diminish the statistical efficiency of RDD. In the phase~II trial of sorafenib \citep{Ratain2006}, only patients who achieved stable disease (less than $\pm$25\% change in tumor size) after initial treatment with sorafenib were eligible to enter the randomized part. In a commentary, \cite{Sonpavde2006} questioned whether such enrichment criteria would result in a relatively homogeneous population for a randomized comparison. In response, \cite{Ratain2006b} emphasized that the conducted phase~II RDD trial was still exploratory (and yet better than a single arm trial alternative), and informed the subsequent phase~III placebo-controlled RCT which later provided confirmatory evidence of sorafenib efficacy in RCC and its subsequent approval by the US FDA.  Another potentially elusive point of the population attribute in RDD is the existence of the target population of patients to whom the results of the RDD trial should generalize. In the phase~II trial of sorafenib \citep{Ratain2006}, such a population did not exist at the time when the trial was conducted as the drug was still experimental. In the phase~III trial of avelumab \citep{Moehler2020}, the ``enriched'' population included patients who did not have progressive disease after treatment with the standard chemotherapy during the open-label period of the study. One can argue that such a population is legitimately defined given the existence of the standard chemotherapy treatment.

Just as the population attribute in RDDs differs from traditional RCTs, so too does the treatment attribute. In standard RCTs, participants are randomized at the outset to receive predefined interventions, typically across a broad, all-comer population. In contrast, RDDs include an initial open-label run-in phase where all participants receive the experimental treatment. Only those who meet specific criteria -- usually based on response, tolerance, or adherence -- are subsequently randomized to either continue the treatment or switch to a control condition. As a result, the treatment attribute in RDDs reflects treatment maintenance, not treatment initiation. Consequently, any causal inference pertains only to a conditional responder population, not the broader target population. This distinction complicates both the interpretation and the generalizability of the treatment effects. Furthermore, ambiguity may arise regarding what aspect of treatment the estimand is intended to capture: does it reflect the entire treatment journey, or only the maintenance phase following initial benefit? It is therefore essential to clearly define the treatment condition -- whether it represents ``continued active treatment'' versus ``withdrawal or placebo after initial benefit,'' -- and specify whether the estimand is intended to assess maintenance efficacy or the overall value of the treatment from initiation. In a traditional RDD, the primary focus is typically on the former:  maintenance efficacy. In the phase~II trial example \citep{Ratain2006}, all study participants received sorafenib during the open-label part of the trial, and only those with stable disease were randomized to either continue with sorafenib or switch to placebo. In this case, placebo represents the intervention whose effect is being evaluated. However, the research hypothesis aimed to test the superiority of continued sorafenib therapy over placebo, which contrasts with the conventional RCT logic where a superiority hypothesis typically compares an intervention against a control. In the phase~III trial example \citep{Moehler2020}, avelumab was the experimental treatment tested for its superiority as maintenance therapy against standard chemotherapy in the randomized part of the study. This approach aligns with the conventional RCT design.

Regarding the endpoint attribute, a potential issue with RDD arises when the response to treatment in the open-label stage is used as a predictor of response in the randomized stage. \cite{Ghaemi2017} argued that the only way to avoid this tautology is to ensure the predictor and the outcome are different. For example, in a psychiatry study, acute treatment response (treatment of a current episode) can be used as a predictor of maintenance treatment response (prevention of future new episodes). In clinical oncology, primary endpoints can be binary or time-to-event. In the phase~II trial example \citep{Ratain2006}, the primary endpoint was the disease progression status (Yes/No) at 12 weeks following random assignment, i.e., 24 weeks after study entry. This was similar, but not identical to the enrichment endpoint (``stable disease,'' as determined by the change of less than 25\% in tumor size at the end of the 12-week open-label period). In contrast, the phase~III JAVELIN Gastric 100 trial \citep{Moehler2020} used overall survival (time from randomization to death from any cause) as the primary endpoint, which was distinctly different from the binary enrichment predictor of being in a state of ``disease control'' by end of the 12-week induction period. Importantly, the choice of follow-up duration in the randomized period impacts the number of events/censored observations and the statistical power of time-to-event analyses.

The population-level summary refers to the precise description of the quantity or effect that the study aims to estimate across the target population. A hallmark of RDD is that the definition of the population is data-driven -- the ``enriched'' population is selected based on the results of the open-label part of the study. In the phase~II trial of sorafenib \citep{Ratain2006}, the population-level summary was the proportion of patients with advanced RCC (among those who achieved stable disease after first 12 weeks of sorafenib treatment) who remained progression-free after 12 weeks following their randomized assignment.  In the phase~III trial of avelumab \citep{Moehler2020}, the population-level summary was median OS in each randomized treatment group and the (stratified) hazard ratio from a Cox proportional hazards model as a measure of the group difference. Therefore, once the target population is defined, specifying the population summary measure for RDD should be similar to the traditional RCT. 

According to E9(R1), intercurrent events (ICEs) are defined as events occurring after treatment assignment that affect the interpretation or existence of the outcome measure. While RDDs may be perceived to have different ICE profiles compared to conventional RCTs, this distinction largely disappears when the estimand is anchored at the point of randomization. In RDDs, once treatment assignment begins at the start of the randomized discontinuation phase, ICEs and their handling strategies are conceptually and operationally similar to those in parallel-arm RCTs. The perceived differences typically stem from events occurring during the pre-randomization enrichment phase -- such as early progression, toxicity, or treatment discontinuation. However, these are not considered ICEs under the estimand framework if the estimand is defined from the point of randomization and the population (e.g., ``enriched'' vs. overall) and  treatment (e.g.,  maintenance vs. initiation) attributes are clearly specified. These pre-randomization events affect study design and population selection, but they do not affect the treatment effect being estimated -- unless the estimand is explicitly defined to reflect the entire treatment journey, starting from initial therapy. In that case, such events may indeed qualify as ICEs. Overall, estimand strategies in RDDs require the same level of rigor in identifying and handling ICEs as in traditional trial designs, provided the estimand is clearly anchored to the randomized phase. For example, in our considered RDD case studies \citep{Ratain2006, Moehler2020}, premature treatment discontinuation -- with or without subsequent follow-up therapy -- was identified as a key ICE.

\section{Discussion}\label{Sec4}

Since the release of the ICH E9(R1) in 2019, the estimand framework has been increasingly useful in clinical trials \citep{Fierenz2024, Kahan2024}. While this framework is primarily focused on RCTs, its principles are also applicable to any clinical study in which a treatment effect is estimated. In this paper, we considered an application of the estimand framework to the randomized discontinuation design (RDD) in oncology using two published examples -- a phase~II trial of orafenib \citep{Ratain2006}, and a phase III JAVELIN Gastric 100 trial of avelumab \citep{Moehler2020} -- that implemented RDD without considering estimands as per E9(R1). Our work provides some new perspectives on RDD in the context of E9(R1), highlighting some similarities and differences between the RDD and the traditional RCT.

RDD is an adaptive design, classified as a predictive enrichment strategy \citep{FDA2019} that can be used to validate treatment efficacy in a targeted subpopulation of patients who initially respond to treatment. Several recent papers attempted to link the estimand framework with adaptive clinical trials \citep{Okwuokenye2019} and complex innovative designs \citep{Collignon2022}. However, to the best of our knowledge, our current paper is the first one to describe estimands for RDD in oncology trials. 

We found that RDD has several unique features that make the corresponding estimands distinct from those in a traditional RCT. In particular, the definition of a target population in RDD is data-driven -- it includes trial participants whose disease is controlled (based on pre-specified criteria) after initial treatment with the experimental drug, as in the example of the phase~II sorafenib trial \citep{Ratain2006}, or the standard chemotherapy, as in the example of the phase~III avelumab trial \citep{Moehler2020}. Furthermore, the use of a placebo as an intervention for the randomized part of the study and the use of a binary primary endpoint that is closely related to the binary enrichment endpoint adds to the complexity of defining the estimand for a phase~II RDD. 
As regards ICEs, their definition and corresponding handling strategies are, overall, consistent between RDDs and traditional RCTs. Given some unique features of RDDs, it is crucial to ensure that the description of the ICEs aligns with the study objectives, endpoints, and treatment arms during the maintenance phase of the study (i.e., post-randomization). 

An additional important point that merits consideration is the scope of data for statistical analysis. In RDD, typically only data from the randomized part of the study is used, whereas the open-label part is considered as a screening process. Some approaches to statistical inference utilizing data from both stages of RDD were proposed for binary outcomes \cite{FedorovLiu2014} and time-to-event outcomes (PFS) \cite{Karrison2012}. Defining estimands that take into account data from both stages of RDD is a challenging problem that may be worthy of investigation. 

\section*{Author contributions}
Ayon Mukherjee (AM), Oleksandr Sverdlov (OS), Ngoc-Thuy Ha (NTH), and Yu Deng (YD) initiated the project. The concept and the project outline originated from AM. This was then further developed by the other authors. The structure of the manuscript was defined by OS. All authors contributed to the manuscript drafting and read and approved the final manuscript. 

\section*{Acknowledgments}
This work is a result of a cross-industry collaboration between the American Statistical Association Biopharmaceutical Section (ASA BIOP) Randomization Working Group (RWG) and the European Federation for Statisticians in Pharmaceutical Industry (EFSPI) Estimand Working Group (EIWG). AM and OS are a part of the RWG, and NTH and YD are a part of EIWG. The authors would like to thank Newcastle University, Novartis, Merck KGaA, and Genentech Inc. for providing resources in support of completing this research work. 

\section*{Financial disclosure}
The authors received no specific funding for this work.

\section*{Conflict of interest}
The authors declare that they have no potential conflict of interest.

\section*{Data Availability Statement}
Data sharing not applicable to this article as no datasets were generated or analyzed during the current study.

\section*{ORCID}
Ayon Mukherjee\; 
\url{https://orcid.org/0000-0001-5461-1737} \\
Oleksandr Sverdlov\; 
\url{https://orcid.org/0000-0002-1626-2588} \\
Ngoc-Thuy Ha\; 
\url{https://orcid.org/0000-0002-4846-7238} \\
Yu Deng\; 
\url{https://orcid.org/0009-0004-4554-499X} \\

\bibliographystyle{unsrtnat}
\bibliography{template} 

\begin{thebibliography}{23}
\providecommand{\natexlab}[1]{#1}
\providecommand{\url}[1]{\texttt{#1}}
\expandafter\ifx\csname urlstyle\endcsname\relax
  \providecommand{\doi}[1]{doi: #1}\else
  \providecommand{\doi}{doi: \begingroup \urlstyle{rm}\Url}\fi

\bibitem[Amery and Dony(1975)]{AmeryDony1975}
W.~Amery and J.~Dony.
\newblock Clinical trial design avoiding undue placebo treatment.
\newblock \emph{The Journal of Clinical Pharmacology}, 15:\penalty0 674--679, 1975.
\newblock \doi{10.1002/j.1552-4604.1975.tb05919.x}.

\bibitem[Kopec et~al.(1993)Kopec, Abrahamowicz, and Esdaile]{Kopec1993}
J.~A. Kopec, M.~Abrahamowicz, and J.~M. Esdaile.
\newblock Randomized discontinuation trials: utility and efficiency.
\newblock \emph{Journal of Clinical Epidemiology}, 46(9):\penalty0 959--971, 1993.
\newblock \doi{10.1016/0895-4356(93)90163-u}.

\bibitem[{Food~and~Drug~Administration}(2019)]{FDA2019}
{Food~and~Drug~Administration}.
\newblock {Enrichment Strategies for Clinical Trials to Support Approval of Human Drugs and Biological Products. Guidance for Industry}, 2019.
\newblock \href{https://www.fda.gov/media/121320/download}{https://www.fda.gov/media/121320/download}.

\bibitem[Rosner et~al.(2002)Rosner, Stadler, and Ratain]{Rosner2002}
G.~L. Rosner, W.~Stadler, and M.~J. Ratain.
\newblock Randomized discontinuation design: application to cytostatic antineoplastic agents.
\newblock \emph{Journal of Clinical Oncology}, 20(22):\penalty0 4478--4484, 2002.
\newblock \doi{10.1200/JCO.2002.11.126}.

\bibitem[Trippa et~al.(2012)Trippa, Rosner, and Müller]{Trippa2012}
L.~Trippa, G.~L. Rosner, and P.~Müller.
\newblock Bayesian enrichment strategies for randomized discontinuation trials.
\newblock \emph{Biometrics}, 68\penalty0 (1):\penalty0 203--211, 2012.
\newblock \doi{10.1111/j.1541-0420.2011.01623.x}.

\bibitem[Ratain et~al.(2006)Ratain, Eisen, Stadler, Flaherty, Kaye, Rosner, Gore, Desai, Patnaik, Xiong, Rowinsky, Abbruzzese, Xia, Simantov, Schwartz, and O'Dwyer]{Ratain2006}
M.~J. Ratain, T.~Eisen, W.~M. Stadler, K.~T. Flaherty, S.~B. Kaye, G.~L. Rosner, M.~Gore, A.~A. Desai, A.~Patnaik, H.~Q. Xiong, E.~Rowinsky, J.~L. Abbruzzese, C.~Xia, R.~Simantov, B.~Schwartz, and P.~J. O'Dwyer.
\newblock Phase {II} placebo-controlled randomized discontinuation trial of sorafenib in patients with metastatic renal cell carcinoma.
\newblock \emph{Journal of Clinical Oncology}, 24(16):\penalty0 2505--2512, 2006.
\newblock \doi{10.1200/JCO.2005.03.6723}.

\bibitem[Nosov et~al.(2012)Nosov, Esteves, Lipatov, Lyulko, Anischenko, Chacko, Doval, Strahs, Slichenmyer, and Bhargava]{Nosov2012}
D.~A. Nosov, B.~Esteves, O.~N. Lipatov, A.~A. Lyulko, A.~A. Anischenko, R.~T. Chacko, D.~C. Doval, A.~Strahs, W.~J. Slichenmyer, and P.~Bhargava.
\newblock {Antitumor activity and safety of tivozanib (AV-951) in a phase II randomized discontinuation trial in patients with renal cell carcinoma}.
\newblock \emph{Journal of Clinical Oncology}, 30\penalty0 (14):\penalty0 1678--1685, 2012.
\newblock \doi{10.1200/JCO.2011.35.3524}.

\bibitem[Smith et~al.(2013)Smith, Smith, Sweeney, Elfiky, Logothetis, Corn, Vogelzang, Small, Harzstark, Gordon, Vaishampayan, Haas, Spira, Lara, Lin, Srinivas, Sella, Schöffski, Scheffold, Weitzman, and Hussain]{Smith2013}
D.~C. Smith, M.~R. Smith, C.~Sweeney, A.~A. Elfiky, C.~Logothetis, P.~G. Corn, N.~J. Vogelzang, E.~J. Small, A.~L. Harzstark, M.~S. Gordon, U.~N. Vaishampayan, N.~B. Haas, A.~I. Spira, P.~N. Lara, C.~C. Lin, S.~Srinivas, A.~Sella, P.~Schöffski, C.~Scheffold, A.~L. Weitzman, and M.~Hussain.
\newblock Cabozantinib in patients with advanced prostate cancer: results of a phase {II} randomized discontinuation trial.
\newblock \emph{Journal of Clinical Oncology}, 31\penalty0 (4):\penalty0 412--419, 2013.
\newblock \doi{10.1200/JCO.2012.45.0494}.

\bibitem[Harding et~al.(2018)Harding, Amato, Stuy, Longo, Burton, Posner, Weng, Merilainen, Gu, Jiang, Vockley, and {PRISM-2~Investigators}]{Harding2018}
C.~O. Harding, R.~S. Amato, M.~Stuy, N.~Longo, B.~K. Burton, J.~Posner, H.~H. Weng, M.~Merilainen, Z.~Gu, J.~Jiang, J.~Vockley, and {PRISM-2~Investigators}.
\newblock {Pegvaliase for the treatment of phenylketonuria: A pivotal, double-blind randomized discontinuation Phase 3 clinical trial}.
\newblock \emph{Molecular Genetics and Metabolism}, 124\penalty0 (1):\penalty0 20--26, 2018.
\newblock \doi{10.1016/j.ymgme.2018.03.003}.

\bibitem[Moehler et~al.(2021)Moehler, Dvorkin, Boku, Özgüroğlu, Ryu, Muntean, Lonardi, Nechaeva, Bragagnoli, Coşkun, {Cubillo~Gracian}, Takano, Wong, Safran, Vaccaro, Wainberg, Silver, Xiong, Hong, Taieb, and Bang]{Moehler2020}
M.~Moehler, M.~Dvorkin, N.~Boku, M.~Özgüroğlu, M.~H. Ryu, A.~S. Muntean, S.~Lonardi, M.~Nechaeva, A.~C. Bragagnoli, H.~S. Coşkun, A.~{Cubillo~Gracian}, T.~Takano, R.~Wong, H.~Safran, G.~M. Vaccaro, Z.~A. Wainberg, M.~R. Silver, H.~Xiong, J.~Hong, J.~Taieb, and Y.~J. Bang.
\newblock {Phase III trial of avelumab maintenance after first-line induction chemotherapy versus continuation of chemotherapy in patients with gastric cancers: Results from JAVELIN Gastric 100}.
\newblock \emph{Journal of Clinical Oncology}, 39\penalty0 (9):\penalty0 966--977, 2021.
\newblock \doi{10.1200/JCO.20.00892}.

\bibitem[{Food~and~Drug~Administration}(2021)]{FDA2021}
{Food~and~Drug~Administration}.
\newblock {E9~(R1) Statistical Principles for Clinical Trials: Addendum: Estimands and Sensitivity Analysis in Clinical Trials}, 2021.
\newblock \href{https://www.fda.gov/media/148473/download}{https://www.fda.gov/media/148473/download}.

\bibitem[Escudier et~al.(2007)Escudier, Eisen, Stadler, Szczylik, Oudard, Siebels, Negrier, Chevreau, Solska, Desai, Rolland, Demkow, Hutson, Gore, Freeman, Schwartz, Shan, Simantov, Bukowski, and {TARGET~Study~Group}]{Escudier2007}
B.~Escudier, T.~Eisen, W.~M. Stadler, C.~Szczylik, S.~Oudard, M.~Siebels, S.~Negrier, C.~Chevreau, E.~Solska, A.~A. Desai, F.~Rolland, T.~Demkow, T.~E. Hutson, M.~Gore, S.~Freeman, B.~Schwartz, M.~Shan, R.~Simantov, R.~M. Bukowski, and {TARGET~Study~Group}.
\newblock Sorafenib in advanced clear-cell renal-cell carcinoma.
\newblock \emph{New England Journal of Medicine}, 356\penalty0 (2):\penalty0 125--134, 2007.
\newblock \doi{10.1056/NEJMoa060655}.

\bibitem[Mütze et~al.(2025)Mütze, Bell, Englert, Hougaard, Jackson, Lanius, and Ravn]{Mutze2025}
T.~Mütze, J.~Bell, S.~Englert, P.~Hougaard, D.~Jackson, V.~Lanius, and H.~Ravn.
\newblock Principles for defining estimands in clinical trials---{A} proposal.
\newblock \emph{Pharmaceutical Statistics}, 24:\penalty0 e2432, 2025.
\newblock \doi{10.1002/pst.2432}.

\bibitem[Fedorov(2022)]{Fedorov2022}
V.~V. Fedorov.
\newblock Randomized discontinuation trials.
\newblock In Steven Piantadosi and Curtis~L. Meinert, editors, \emph{Principles and Practice of Clinical Trials}, pages 1439--1453. Springer Nature Switzerland AG, 2022.
\newblock \doi{10.1007/978-3-319-52636-2}.

\bibitem[Sonpavde et~al.(2006)Sonpavde, Hutson, Galsky, and Berry]{Sonpavde2006}
G.~Sonpavde, T.~E. Hutson, M.~D. Galsky, and W.~R. Berry.
\newblock Problems with the randomized discontinuation design.
\newblock \emph{Journal of Clinical Oncology}, 24\penalty0 (28):\penalty0 4669--4670, 2006.
\newblock \doi{10.1200/JCO.2006.08.102}.

\bibitem[Ratain(2006)]{Ratain2006b}
M.~J. Ratain.
\newblock {In reply to the letter by Sonpavde et al.}
\newblock \emph{Journal of Clinical Oncology}, 24\penalty0 (28):\penalty0 4670--4671, 2006.
\newblock \doi{10.1200/JCO.2006.08.4418}.

\bibitem[Ghaemi and Selker(2017)]{Ghaemi2017}
S.~N. Ghaemi and H.~P. Selker.
\newblock {Maintenance efficacy designs in psychiatry: Randomized discontinuation trials - enriched but not better}.
\newblock \emph{Journal of Clinical and Translational Science}, 1\penalty0 (3):\penalty0 198--204, 2017.
\newblock \doi{10.1017/cts.2017.2}.

\bibitem[Fierenz and Zapf(2024)]{Fierenz2024}
A.~A. Fierenz and A.~Zapf.
\newblock Current developments of the estimand concept.
\newblock \emph{Pharmaceutical Statistics}, 23:\penalty0 864--869, 2024.
\newblock \doi{10.1002/pst.2395}.

\bibitem[Kahan et~al.(2024)Kahan, Hindley, Edwards, Cro, and Morris]{Kahan2024}
B.C. Kahan, J.~Hindley, M.~Edwards, S.~Cro, and T.P. Morris.
\newblock {The estimands framework: a primer on the ICH E9(R1) addendum}.
\newblock \emph{BMJ}, 384:\penalty0 e076316, 2024.
\newblock \doi{10.1136/bmj-2023-076316}.

\bibitem[Okwuokenye and Peace(2019)]{Okwuokenye2019}
M.~Okwuokenye and K.~E. Peace.
\newblock Adaptive design and the estimand framework.
\newblock \emph{Annals of Biostatistics \& Biometric Applications}, 1\penalty0 (5):\penalty0 1--4, 2019.
\newblock \doi{10.33552/ABBA.2019.01.000524}.

\bibitem[Collignon et~al.(2022)Collignon, Schiel, Burman, Rufibach, Posch, and Bretz]{Collignon2022}
O.~Collignon, A.~Schiel, C.~F. Burman, K.~Rufibach, M.~Posch, and F.~Bretz.
\newblock Estimands and complex innovative designs.
\newblock \emph{Clinical Pharmacology \& Therapeutics}, 112\penalty0 (6):\penalty0 1183--1190, 2022.
\newblock \doi{10.1002/cpt.2575}.

\bibitem[Fedorov and Liu(2014)]{FedorovLiu2014}
V.V. Fedorov and T.~Liu.
\newblock Randomized discontinuation trials with binary outcomes.
\newblock \emph{Journal of Statistical Theory and Practice}, 8:\penalty0 30--45, 2014.
\newblock \doi{10.1080/15598608.2014.840492}.

\bibitem[Karrison et~al.(2012)Karrison, Ratain, W.M., and Rosner]{Karrison2012}
T.G. Karrison, M.J. Ratain, Stadler W.M., and G.L. Rosner.
\newblock Estimation of progression-free survival for all treated patients in the randomized discontinuation trial design.
\newblock \emph{The American Statistician}, 66:\penalty0 155--162, 2012.
\newblock \doi{10.1080/00031305.2012.720900}.

\end{thebibliography}

\end{document}